\documentclass[runningheads]{svmult}

\usepackage{graphicx}  
\usepackage{subeqnar}  
\usepackage{multicol}  
\usepackage{physmubb}  

\begin{document}
\title*{Optimal use of Information for Measuring $M_t$ in
Lepton+jets $t\overline{t}$ Events.}
\toctitle{Optimal use of Information for Measuring $M_t$ in
Lepton+jets $t\overline{t}$ Events.}
%
%
\titlerunning{Measuring $M_t$ in $t\overline{t}$ events.}
%
\author{Juan Estrada, Fermilab
\newline (On behalf of the D\O\ collaboration)}
\authorrunning{Juan Estrada on behalf of the D\O\ Collaboration.}
%
%

\maketitle              

The observation of the top ($t$) quark served as one of the major
confirmations  of the validity of the standard model (SM) of particle
interactions \cite{cdfPRD1,d0PRD2}. Through radiative corrections of the SM,
the mass of the top quark, along with that of the W boson \cite{pdg3}, provide
the best indication for the value of the mass of the hypothesized Higgs
boson \cite{lepEWWG4}. The mass of the W is known to a precision of $< 0.1$\%,
while the uncertainty on the mass of the top quark is at the 4\% level \cite{pdg3}.
Improvements in both measurements are required to limit the range of mass
that the Higgs boson can assume in the SM, and, of course, to check
whether that agrees with expectation. It is therefore important to
develop techniques for extracting the mass of the top quark that can
provide the sharpest values possible.

\section*{Measurement of $M_t$}
We report on a new preliminary measurement of the mass of the top quark from $t \bar t$
data in lepton+jets channels accumulated by the D\O\ experiment in Run-1
of the Tevatron. The luminosity corresponds to 125 events/pb, and this
analysis is based on the same data sample that was used to extract the
mass of the top quark in our previous publication \cite{massPRD5}. Information
pertaining to the detector and to the older analysis can be found in
Refs. \cite{d0NIM6} and \cite{massPRD5}, respectively.

After offline selections on lepton transverse energies ($E_T > 20$ GeV) and
angles ($|\eta_{\mu}| < 1.7$, and $|\eta_{e}| < 2.0$, for muon and electron
channels, respectively), on jet transverse energies ($E_T > 15$ GeV) and
angles ($|\eta| < 2.0$), imbalance in transverse momentum (missing-$E_T >20$
GeV), and after applying several less important criteria \cite{massPRD5}, the
event sample consisted of 91 events with one isolated lepton and four or
more jets. (Unlike the previous analysis, we do not distinguish between
events that have or lack a muon associated with one of the jets, signifying
the possible presence of a $b$-quark jet in the final state.) The new
analysis involves a comparison of the data with a leading-order matrix
element for the production and decay process, and to minimize the effect
of higher-order corrections we therefore restrict the study to events
containing only four jets. This requirement reduces data sample to 71
events.

\begin{figure}
\begin{center}
\setlength{\unitlength}{1mm}
\begin{picture}(80,80)(0,0)
\put(0,0){\includegraphics[width=.6\textwidth]{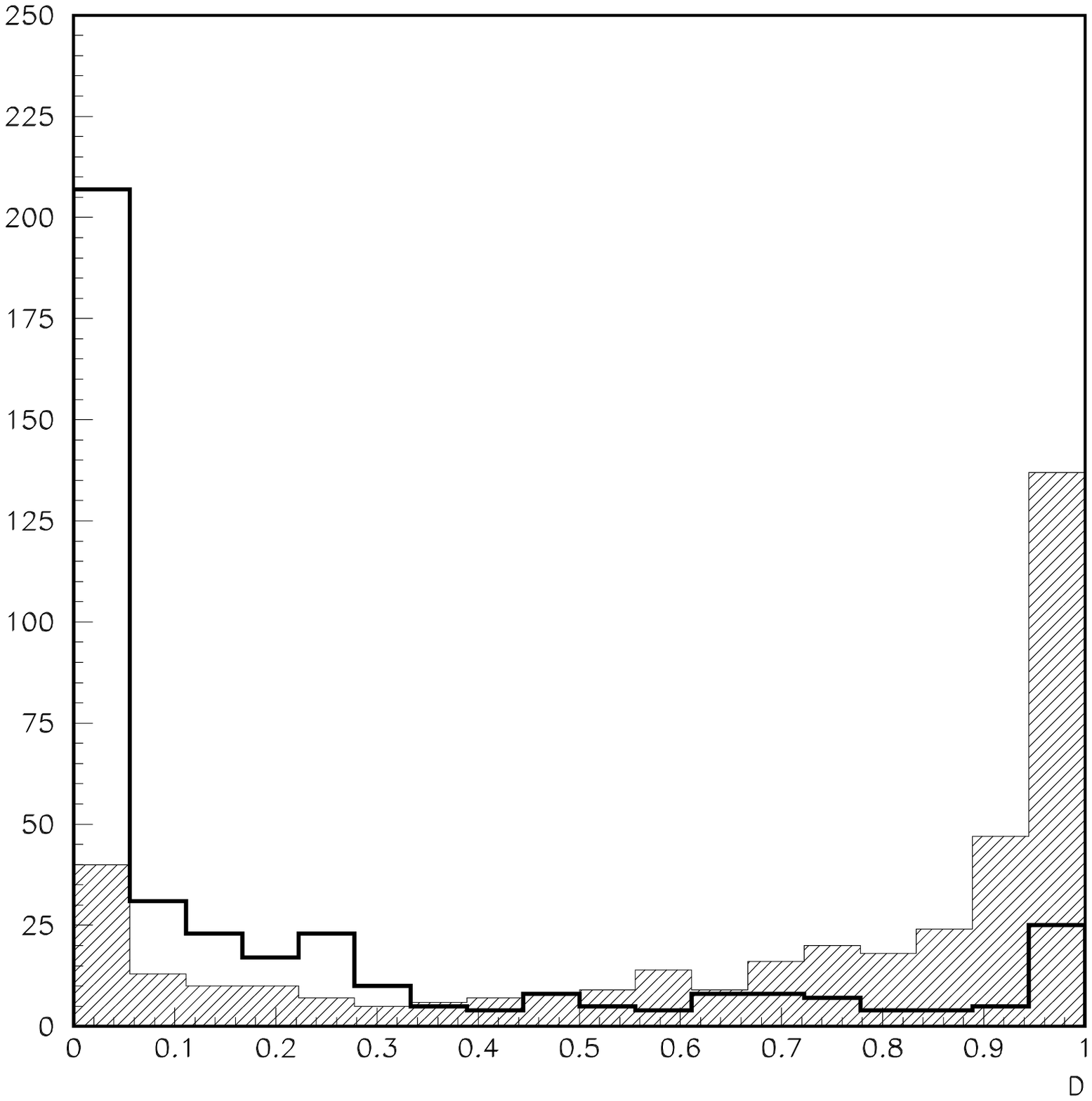}}
\put(10,35){\includegraphics[width=.5\textwidth]{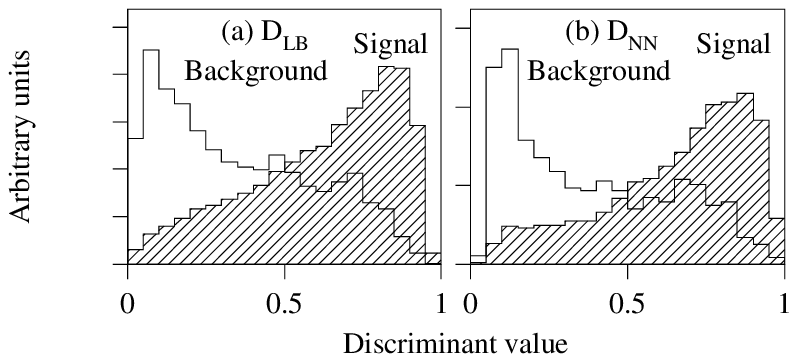}}
\end{picture}
\end{center}
\caption[]{Discriminator $D=P_{signal}/(P_{signal}+P_{backg.})$ for the
new analysis, and (inset) for the two versions of the
analysis included in the previous D\O\ publication \cite{massPRD5}.}
\label{fig:discr}
\end{figure}

In the previous analysis, the four jets with highest $E_T$ were assumed
to represent the four quarks in the events (two $b$ quarks from the
initial $t\rightarrow W+b$ decays, and the $q' \bar q$ decays of one
of the $W$ bosons). These, along with the lepton and the missing $\nu$
(from the decay of the other $W$ boson) were fitted to the kinematic
hypothesis $p \bar p \rightarrow t \bar t \rightarrow W W b \bar b$,
subject to the constraints of overall momentum-energy
conservation, the known mass of the $W$ boson, and the fact that
the unknown mass of the top quark ($M_t$) had to be identical for
the top and antitop quarks in the event. With 12 ways to permute
the jets, there were 12 possible fits, and the solution with the
lowest $\chi^2$ (but $<$ 10) was chosen as the best hypothesis,
thereby defining the fitted mass $m_{fit}$ for the event, as well
as the longitudinal momentum of the $\nu$. The same procedure was
used for generating templates in variables of interest, as a function
of input mass of the top quark. This was based on the Monte Carlo (MC)
HERWIG program \cite{herwig7}, which was used to generate events that were
passed through full event-reconstruction in the detector \cite{d0geant8}.
The background, which consisted mainly of all-jet production (20\%)
and $W$+jets (80\%) was also processed in a similar manner. The
background from all-jet production was based on data, and the
background from $W$+jets was based on events generated with
VECBOS \cite{vecbos9}. A discriminant was formed based on differences
in distributions in four variables predicted for $t \bar t$ signal
and background, and this defined the probability that any event
represented signal as opposed to background. A probability density was
defined as a function of the discriminant ($D$) and $m_{fit}$,
and a comparison of data and MC via a likelihood was used to
determine the most likely mass of the top quark.

The new analysis is also based on a likelihood, but this likelihood
is a function of all measured variables in the event, with the
exception of the unclustered energy in the calorimeter that is used
to define the missing $E_T$. Our method is similar to that suggested
for $t \bar t$ dilepton decay channels \cite{dgk10}, and used in previous
mass analyses of dilepton events \cite{dilep11}. We define a probability
for background purely in terms of the matrix element contained in
VECBOS, and for signal in terms of the leading-order matrix element
for $t \bar t$ production and decay. This is convoluted with a
transfer function that relates objects at the parton level to the
objects observed (fully reconstructed) in the detector. We assume
that the observed electrons correspond to the produced electrons,
and that the muons are smeared with their known resolution. The
angles of the jets are assumed to reflect the angles of the partons
in the final state, and we ignore any transverse momentum
for the incident partons. We use a large MC sample of $t \bar t$
events (generated with masses between 140$-$200 GeV in HERWIG,
and processed through the D\O\ detector-simulation package) to
determine a set of ten parameters that correlate any jet $E_T$ with
its parton value. The parameters used for $b$ quarks are
different than for the lighter quarks, and there are therefore
20 parameters in all.

We can write the probability for $t \bar t$ production in terms of
the following elements:
\begin{equation}
P_{t \bar t}= \int d\rho_1 dm_1^2 dM_1^2 dm_2^2 dM_2^2 \sum_{\mbox{perm.},\nu}
               |M|^2 \frac{f(q_1)f(q_2)}{|q_1||q_2|} \Phi_6 W(x,y)
\end{equation}
where the sum is over all 12 permutations of the jets, and all
possible values of $\nu$ momenta. $|M|^2$ is the matrix element for the process,
$f(q_1)$ and $f(q_2)$ are the parton distribution
function (PDF) for the incident particles, $\Phi_6$
is the phase-space factor for the 6-object final state
and $W(x,y)$ correspond to a function
that parameterizes the mapping between parton level quantities
$y$ and detector measurements $x$.
With two incident parton energies (we take these partons to be quarks,
and ignore the $\approx 10\% $ contribution from gluon fusion), and six
objects in the final state, the integrations over the essentially fifteen
sharp variables (3 components of lepton momentum, eight jet angles,
and four $\delta$-functions representing energy-momentum conservation),
leaves five integrals that must be performed to obtain the probability
that any event represents $t \bar t$ production for some specified
value of top mass $m_t$. Four of the variables chosen for the remaining
integrations, namely the masses of the $W$ bosons and of the top quarks
in the event, are economical in CPU time, because the value of $|M|^2$
is essentially negligible except at the peaks of the four Breit-Wigner
terms in the matrix element.


\begin{figure}
\begin{center}
\includegraphics[width=.6\textwidth]{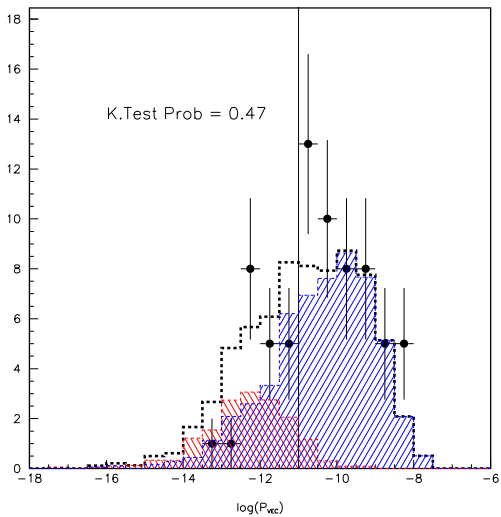}
\end{center}
\caption[]{Distribution for background probability calculated for
the 71 4-jets $t \bar t$ candidates (data points). Only the events to the left
of the vertical line ($\ln[P_{backg}]<-11,0$) are considered
for further analysis. The data is compared with
the expected results from MC-simulated samples (dashed histrogram) formed
with a mixture of 16 signal (left-hatched histrogram) and 55
background (right-hatched histogram) events.}
\label{fig:backgprob}
\end{figure}

Introducing an analogous expression for the background, the likelihood
as a function of $m_t$ can be written as:
\begin{eqnarray}
-ln L(\alpha) = &-& \sum_{i=1}^N \ln[c_1 P_{t \bar t}(x_i; \alpha) + c_2 P_{bkg}(x_i)] \nonumber \\
                &+& N\int A(x) [c_1 P_{t \bar t}(x; \alpha) + c_2 P_{bkg}(x)] dx
\end{eqnarray}
where $A(x)$ is the acceptance of the detector in terms of the parton
variables (and contains the transfer function), and the sum over the
probabilities for the individual events reflects the product of the
probabilities for final sample of $N$ of events. The best values of
$\alpha$, representing $m_t$, and the parameters $c_i$, are all defined by
the most probable value of the likelihood.

The main difference between this method and the previous analysis is
that each event now has its individual probability as a function of the
mass parameter. This probability, reflecting both signal and background,
depends on all measured variables in the event (excepting unclustered
energy), with well-measured events contributing more sharply to the
extraction of the mass of the top quark than those poorly measured.
The fact that so much of the information in each event is retained in
the probability provides a far better separation of signal and
background. Figure \ref{fig:discr} displays the effective difference in the
discriminating power for the parameter $D$ in the current and in the
previous analyses. In all these analyses, this parameter is not used to
select regions of high signal-probability, but the comparison is
instructive in showing the greater power of the present method.

\begin{figure}
\begin{center}
\includegraphics[width=.9\textwidth]{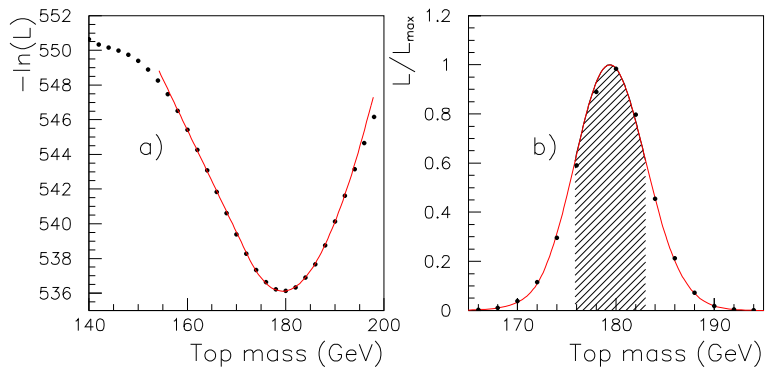}
\end{center}
\caption[]{a) Negative of the log of the
likelihood as a function of the mass of the top quark for
the 22 $t \bar t$ candidates in our final sample. b)
Probability distribution determined from the likelihood, with the
hatched area corresponding to the 68\% probability interval,
from which we determine $M_t = 179.9 \pm 3.6$ GeV. A 0.5 GeV
upward bias correction has been
included based on the results of MC ensemble tests.}
\label{fig:likelihood}
\end{figure}

Studies on samples of HERWIG MC events used in the previous analysis
indicate that the expected uncertainty using the new method would
be capable of yielding almost a factor of two reduction in the
uncertainty of the extracted top mass. However, the method appears
to be sensitive to a systematic shift of the mass (of about 2 GeV)
as a function of the {\it a priori} sample purity, whose true value cannot
be determined in the analysis. To minimize the impact of this limiting
systematic uncertainty, we proceed as follows. Figure \ref{fig:backgprob},
shows the probability that each event
corresponds to background (VECBOS), and we compare that to a
sample of 71 MC events corresponding to the mixture of 16 signal and
55 background events found in the data sample. From MC
studies we fibd that eliminating from further consideration in the analysis
those events that have a high probability of being background, reduces
markedly the dependence on the extracted mass on sample purity. We consequently retain
for further study only those events that have a poor probability of
being background. This selection is shown in Fig \ref{fig:backgprob}, and its
imposition leaves a sample of only 22 events, 12 of which are
signal and 10 background, after the recalculation of the new
minimum in the likelihood, the result of which is shown in Fig. \ref{fig:likelihood}

\begin{figure}
\begin{center}
\includegraphics[width=.9\textwidth]{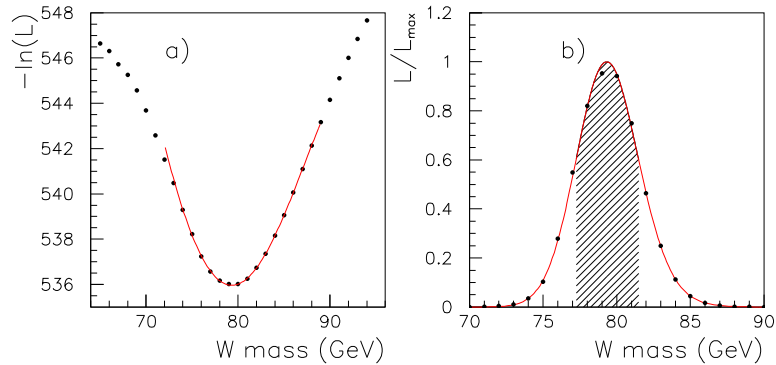}
\end{center}
\caption[]{a) Negative of the log of the
likelihood as a function of the mass of the $W$ boson for
the 22 $t \bar t$ candidates in our final sample. b)
Probability distribution determined from the likelihood, with
the hatched area corresponding to the 68\%.
probability interval.}
\label{fig:likemw}
\end{figure}

Another aspect of this analysis is that
it provides a natural method for measuring
any other parameter in the $t \bar t$ differential
cross section. As an example we, show in Fig.~\ref{fig:likemw}
the likelihood as a function of the mass of the $W$ boson ($M_W$) for
the same sample of events. This suggests the very
interesting possibility of checking the jet energy scale (JES) in this events,
which corresponds to the largest contribution to the systematic uncertainty in the current
and previous analyses. The next step in this analysis is, in fact, to
utilize the information on JES contained in these
events. However, in this preliminary result, we only estimate
this uncertainty using the previous analysis \cite{massPRD5},
where the JES calibration from $\gamma+jet$ events was used to match
the energy scale in the experiment to that in the MC simulation.
Here we estimate the uncertainty in $M_t$ coming from JES as follows.
The analysis is performed before and
after applying this correction, and the difference defined as
the systematic uncertainty due to JES. Table \ref{tb:sys}
summarizes all the systematic uncertainties in this preliminary
result.

\section*{Conclusion}

We have presented a new preliminary measurement of the mass
of the top quark using a method that compares each individual
event with the differential cross section for $t \bar t$
production and decay. We obtain a significant
improvement over the statistical uncertainty
of the previous measurement \cite{massPRD5}, that is
equivalent to having a factor of 2.4 more data.
The new preliminary result is:
\begin{equation}
M_t=179.9 \pm 3.6 \mbox{(stat)} \pm 6.0 \mbox{(sys)} \mbox{ GeV}
\end{equation}

The possibility of checking the value of the $W$ mass in the same events
offers the possibility of redoing the largest systematic uncertainty,
of the same events provides a new handle on controlling the largest systematic,
namely the jet energy scale.

\begin{table}
\caption{Table of systematic uncertainties for the measurement of $M_t$}
\begin{center}
\renewcommand{\arraystretch}{1.4}
\setlength\tabcolsep{5pt}
\begin{tabular}{ll}
\hline
\noalign{\smallskip}
 Model for $t \bar t$               & 1.5 GeV \\
 Model for backgound($W$+jets)      & 1.0 GeV \\
 Noise and multiple interactions & 1.3 GeV \\
 Jet energy scale                & 5.6 GeV \\
 Parton distribution function    & 0.2 GeV \\
 Acceptance correction           & 0.5 GeV \\
\hline
Total                            & 6.0 GeV \\
\end{tabular}
\end{center}
\label{tb:sys}
\end{table}

\section*{Aknowledgements}

We thank F. Canelli, G. Gutierrez and T. Ferbel for their important
contributions to this analysis.

%

\end{document}